\documentclass[aps,amsmath,amssymb,superscriptaddress,floatfix,reprint]{revtex4-2}

\usepackage{color}
\usepackage{graphicx}
\usepackage{dcolumn}
\usepackage{bm}
\usepackage{float}
\usepackage{ulem}

\usepackage[utf8]{inputenc}
\usepackage[T1]{fontenc}
\usepackage{mathptmx}
\usepackage{etoolbox}

\makeatletter
\def\@email#1#2{%
 \endgroup
 \patchcmd{\titleblock@produce}
  {\frontmatter@RRAPformat}
  {\frontmatter@RRAPformat{\produce@RRAP{*#1\href{mailto:#2}{#2}}}\frontmatter@RRAPformat}
  {}{}
}%
\makeatother

\begin{document}

\preprint{APS/123-QED}

\title{Nonlinear planar Hall effect from superconducting vortex motion}


\author{Mio Hashimoto}
\affiliation{Department of Basic Science, The University of Tokyo, Meguro, Tokyo 153-8902, Japan}
\author{Takako Konoike}
\affiliation{Research Center for Materials Nanoarchitectonics (MANA), National Institute for Materials Science, 3-13 Sakura, Tsukuba, Ibaraki 305-0003, Japan}
\author{Tomoki Kobayashi}
\affiliation{Department of Basic Science, The University of Tokyo, Meguro, Tokyo 153-8902, Japan}
\author{Shintaro Hoshino}
\affiliation{Department of Physics, Saitama University, Shimo-Okubo, Saitama 338-8570, Japan}%
\author{Takuya Kawada}
\affiliation{Department of Basic Science, The University of Tokyo, Meguro, Tokyo 153-8902, Japan}
\email{takuyakawada@g.ecc.u-tokyo.ac.jp}
\author{Tomoyuki Yokouchi}
\affiliation{Department of Basic Science, The University of Tokyo, Meguro, Tokyo 153-8902, Japan}
\affiliation{RIKEN Center for Emergent Matter Science (CEMS), Wako, Saitama 351-0198, Japan}
\author{Shinya Uji}
\affiliation{Research Center for Materials Nanoarchitectonics (MANA), National Institute for Materials Science, 3-13 Sakura, Tsukuba, Ibaraki 305-0003, Japan}
\author{Atsutaka Maeda}
\affiliation{Department of Basic Science, The University of Tokyo, Meguro, Tokyo 153-8902, Japan}
\author{Yuki Shiomi}
\affiliation{Department of Basic Science, The University of Tokyo, Meguro, Tokyo 153-8902, Japan}

\date{\today}

\begin{abstract}
We report the nonreciprocal charge transport along the longitudinal and transverse directions in the vortex flow regime of FeSe superconducting films. Clear nonreciprocal signals under an inplane magnetic field reveals symmetry breaking at the film surfaces since the crystal structure of FeSe is centrosymmetric. Although the symmetry in such polar superconductors allows the nonreciprocal transverse response under a magnetic field parallel to the electric current, its observation is physically counterintuitive because vortex motion is not expected in this configuration. We propose that thermally excited (anti)vortices due to the two-dimensional nature of FeSe give rise to the nonreciprocal transverse signals when the mirror symmetry is broken by the inplane magnetic field.

\end{abstract}

\maketitle


{\it Introduction-} Research on rectification effects has a very long history in condensed matter physics since their discovery at metal-semiconductor interfaces \cite{bardeen1947surface, mott1939theory}, which establishes the fundamentals of modern electronics, such as p-n junctions. The emergence of rectification effects requires a broken spatial inversion symmetry at the interfaces. The asymmetry in energy levels at the interfaces generates a unidirectional electron flow. However, broken inversion symmetry alone is insufficient to observe the rectification effect in single materials. In addition to breaking the spatial inversion symmetry, breaking the time-reversal symmetry is also necessary. This type of rectification effect is commonly referred to as the nonreciprocal transport effect or magnetochiral anisotropy effect \cite{rikken2001mca}, which has recently been explored in various materials \cite{tokura2018nonreciprocal,atzori2021mca}. 

In nonreciprocal transport effects, the longitudinal and transverse resistivities are different for electric current $I$ flowing to the right ($+I$) and to the left ($-I$). The symmetry argument predicts that for polar systems such as material surfaces and hetero-interfaces, the longitudinal and transverse nonreciprocal signals are expected when the magnetic field ($H$) is applied in the inplane direction perpendicular and parallel to the $I$ direction, respectively. Beyond the symmetry argument, several microscopic mechanisms have been proposed and found to be material-dependent, such as an asymmetric electron scattering \cite{yokouchi2017electrical,aoki2019anomalous,jiang2020electric,ishizuka2020anomalous}, the deformation of the Fermi surface caused by the Zeeman term \cite{ideue2017bulk,he2019nonlinear}, and the Berry curvature effect \cite{yokouchi2023giant}.

Recently, the study of nonlinear transport effects has also gained increasing attention in superconductors, exemplified by the superconducting diode effects \cite{ando2020observation,du2023superconducting,hou2023ubiquitous,anwar2023spontaneous,bauriedl2022supercurrent}. To date, nonreciprocal response along the longitudinal direction (so-called nonreciprocal magnetoresistance) has been reported in various noncentrosymmetric superconductors, for example, in low-symmetry bulk superconductors  \cite{ideue2020giant, wu2022nonreciprocal, du2023superconducting}, and in polar two-dimensional (2D) superconductors \cite{wakatsuki2017nonreciprocal, itahashi2020nonreciprocal, yasuda2019nonreciprocal, masuko2022nonreciprocal}. In most cases, the nonreciprocal magnetoresistance is enhanced in superconductors compared to normal metals \cite{wakatsuki2017nonreciprocal,masuko2022nonreciprocal,itahashi2020nonreciprocal,zhang2020nonreciprocal}. An origin of the enhanced rectification is the contribution of the ratchet motion of vortices in superconductors. Asymmetric pinning potentials for vortices induce nonreciprocal vortex transport in the vortex flow regime, resulting in giant nonreciprocal signals. In addition, a recent theoretical work pointed out that thermally activated vortices and antivotices also contribute to nonreciprocal magnetoresistance in 2D superconductors \cite{hoshino2018nonreciprocal}. 

Although the nonreciprocal longitudinal response has been comprehensively studied in superconductors experimentally and theoretically, the nonreciprocal transport along the Hall direction has been less studied and limited in noncentrosymmetric superconductors \cite{ideue2020giant,itahashi2020prr,itahashi2022giant}. In fact, previous theoretical works predicted the nonreciprocal transverse signal generation even in 2D centrosymmetric superconductors by applying inplane magnetic field to break the mirror symmetry with respect to the magnetic field direction \cite{hoshino2018nonreciprocal,daido2024prr}. To the best of our knowledge, however, no experiments have explicitly demonstrated this effect yet.

In this letter, we study the nonreciprocal transport effects in superconducting FeSe films in detail by measuring current-nonlinear contributions of longitudinal and transverse resistances under magnetic fields. Although the crystal structure of FeSe is centrosymmetric, we observed nonreciprocal transport signals in the vortex flow regime, suggesting that their origin is vortex ratchet motions. The inversion symmetry is only broken at the film interfaces, but clear nonreciprocal signals are observed owing to the topological nature of vortices \cite{lustikova2018vortex}. Remarkably, we found that the nonreciprocal signals are observed in the transverse (Hall) direction when $H$ is applied parallel to $I$. This is highly counterintuitive, since it is believed that the driving force of magnetic-field induced vortex motion is generated in the direction of $\vec{I} \times \vec{H}$ and thus should be zero in the configuration of $H||I$. We tentatively attribute the origin of the nonreciprocal transverse signals to thermally excited (anti)vortices arising from the 2D nature of FeSe. 

\begin{figure}
\includegraphics[width=90mm]{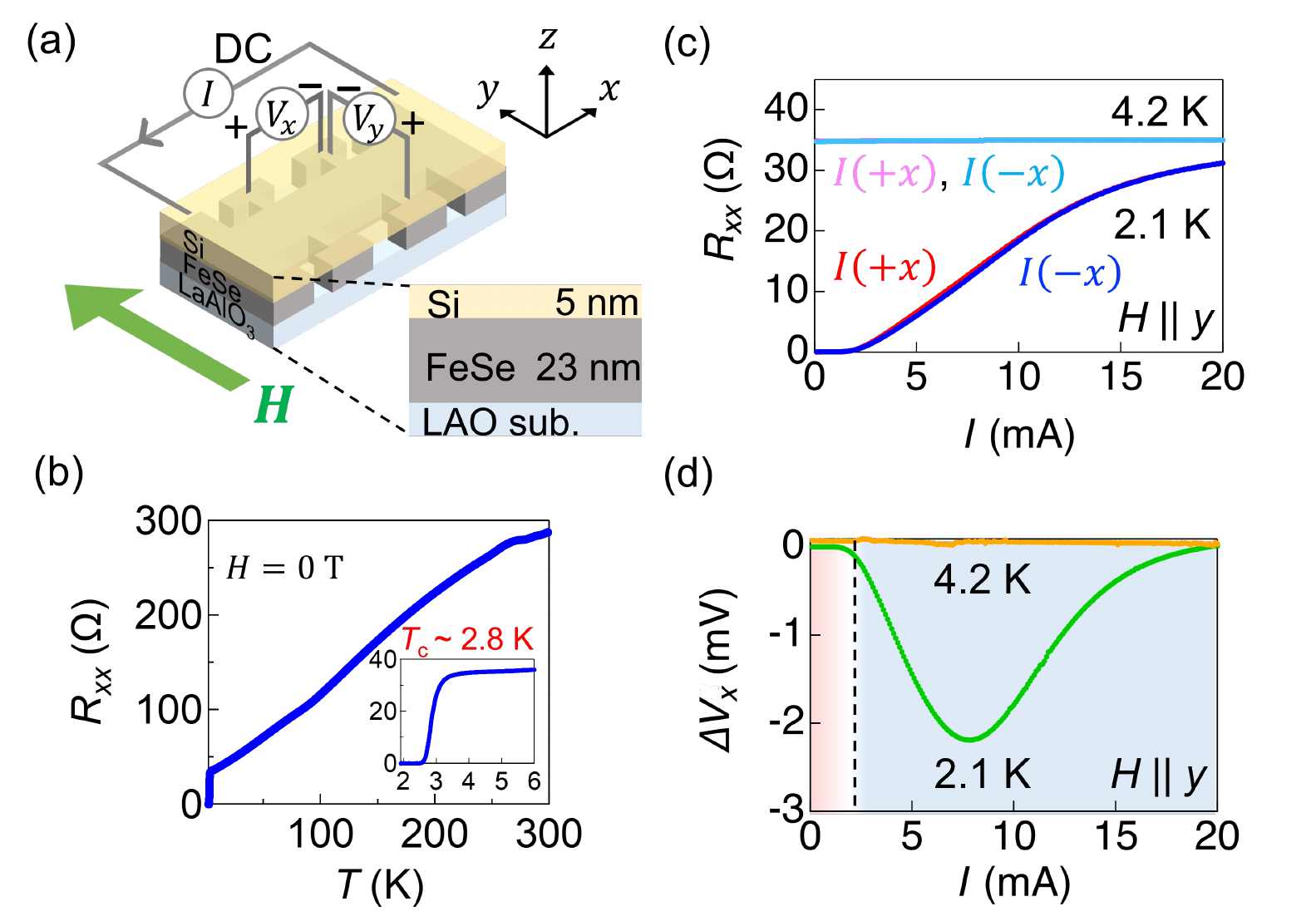}
\caption{\label{fig1} (a) Sample and measurement setup. (b) Temperature ($T$) dependence of resistance ($R_{xx}$). The inset shows a magnified view of the low-$T$ range. (c),(d) Electric-current ($I$) dependence of (c) $R_{xx}$ and (d) nonreciprocal longitudinal voltage $\Delta V_{x}$ at normal state (4.2 K) and superconducting state (2.1 K). The dotted line in (d) represents the position of the depinning current.}
\end{figure}

{\it Methods-} 23-nm-thick FeSe epitaxial films were grown on (001) LaAlO$_3$ (LAO) substrates and capped with amorphous Si by the pulsed laser deposition method \cite{nabeshima2018systematic,hashimotofesesuppl}. The amorphous Si layer is insulating at low temperatures. As shown in Fig. 1(a), the films were deposited in a Hall-bar shape using a metal mask. Transport measurements were conducted using the standard five-terminal method under magnetic fields applied with an Oxford 17 T superconducting magnet. The $H$ directions were rotated with respect to the electric-current direction ($||x$) using a homemade two-axis rotator system. The measurement was mainly performed at 2.1 K below the superconducting transition temperature ($T_{\rm{c}}$); the samples were immersed in superfluid helium with high thermal conductivity, and effects of Joule heating were negligible even when electric current of 20 mA was applied to the films as discussed later.
\par

We used Keithley 2401 Sourcemeter as dc current source and Keithley 2182A Nanovoltmeters for detecting the longitudinal and transverse voltages; whereas lock-in detection methods have been used in most studies of nonreciprocal transport effects, we directly measured the nonreciprocal components of the longitudinal ($||x$) and transverse ($||y$) voltages by taking the difference between the voltages for $+I$ and $-I$ currents at each $H$: $\Delta V_{i} = \{V_{i}(+I) + V_{i}(-I)\}/2$ ($i=x,\ y$). Here, to eliminate background voltage on the transverse voltage $V_y$, we further estimated $\Delta \tilde{V}_{y}(\theta) = \{ \Delta V_{y}(\theta) - \Delta V_{y}(\theta + \pi)\}/2$ in each $H$ direction $\theta(=\alpha,\beta$, and $\gamma$; see Fig. 3 for the definition). By dividing $\Delta V_x$ and $\Delta \tilde{V}_y$ by $I$, the nonreciprocal resistances along the longitudinal and transverse directions, $\Delta R_{xx}$ and $\Delta R_{yx}$, are obtained, respectively.

{\it Results and Discussion-} Figure 1(b) presents the temperature ($T$) dependence of the electrical resistance ($R_{xx}$) of the FeSe film. The sample is metallic and exhibits superconductivity below $T_{\rm{c}} $= 2.8 K, where $T_{\rm{c}}$ is defined as the temperature corresponding to the midpoint of the superconducting transition. Figure 1(c) shows the current ($I$) dependence of electric resistance $R_{xx}$ measured at 2.1 K ($< T_{\rm{c}}$) and 4.2 K ($> T_{\rm{c}}$), both measured under an applied $H$ of 0.6 T along the $+y$ direction. At 4.2 K, $R_{xx}$ is constant with respect to the amplitude of $I$, as expected from the Ohm's law. In contrast, $R_{xx}$ at 2.1 K is almost zero in the small $I$ regime below $\sim 2$ mA, but suddenly increases with a further increase in $I$. Notably, $R_{xx}$ values for $+I$ (shown in red) and $-I$ (blue) are slightly different in the middle $I$ region. Such deviation is not significant at 4.2 K above $T_{\rm{c}}$; $R_{xx}$ for $+I$ (shown in pink) and $-I$ (light blue) is almost the same over the entire $I$ range, showing that nonreciprocal transport of normal-state electrons is negligible because of the centrosymmetric crystal structure.

By taking the difference between the longitudinal voltages measured in $+I$ and $-I$ conditions, $\Delta V_{x}$ is estimated at 2.1 K and 4.2 K, as shown in Fig. 1(d). Whereas $\Delta V_{x}$ is negligibly small in the normal state, a nonzero $\Delta V_{x}$ signal is observed in the superconducting state at 2.1 K. $|\Delta V_{x}|$ exhibits a peak at approximately $8$ mA and decreases with higher $I$ values. This $I$ dependence is attributed to the dynamics of superconducting vortices driven by the applied current. Vortices are pinned by impurities and/or defects inside the superconductor at a low $I$ range, while they are depinned and begin to move as $I$ increases. When the amplitude of $I$ further increases above 8 mA, the superconducting state is gradually destroyed by large electric current ($\sim 10^8$ A/m$^2$) and $\Delta V_{x}$ approaches zero. 

\begin{figure}
\includegraphics[width=90mm]{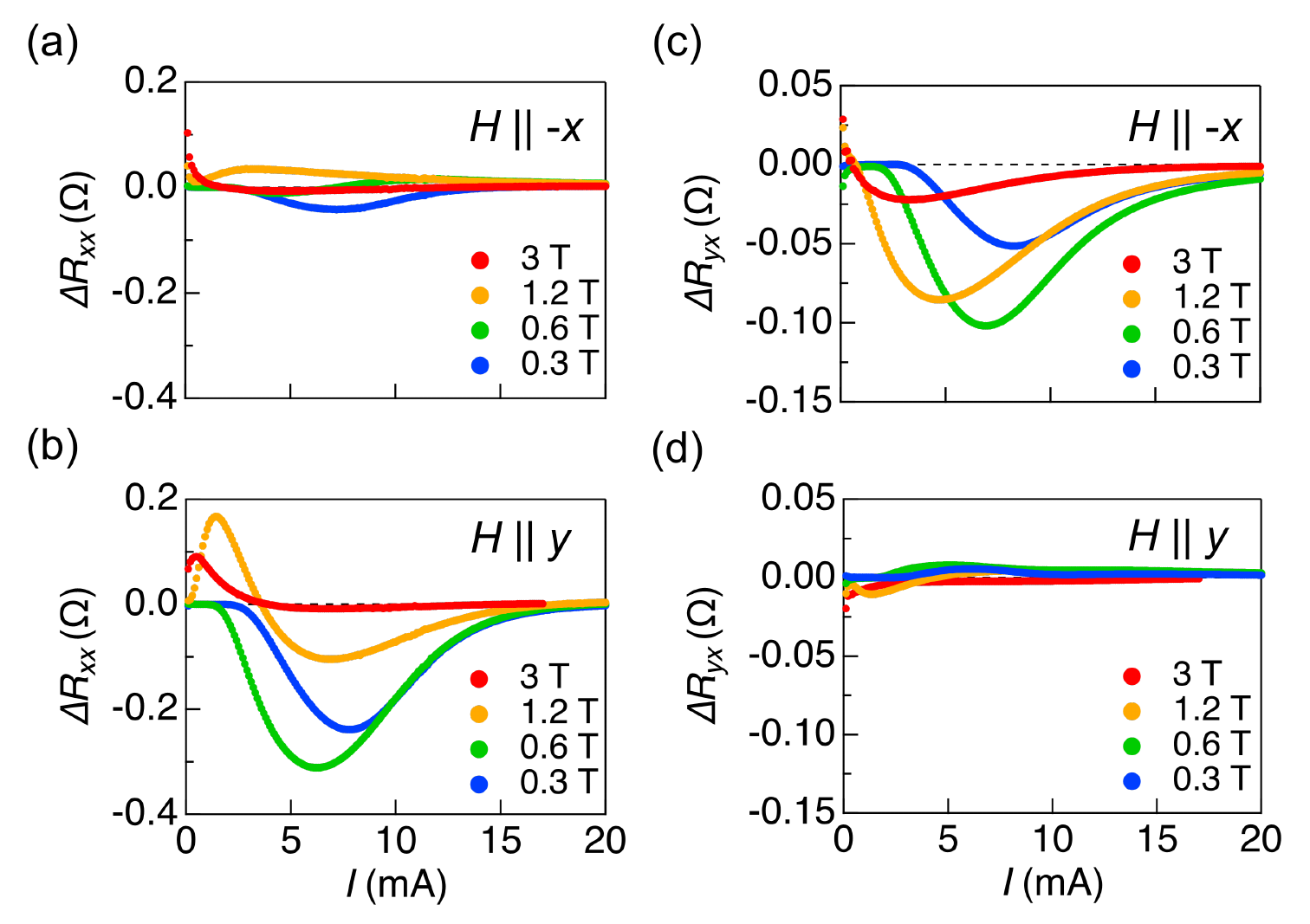}
\caption{\label{fig2} Nonreciprocal longitudinal resistance $\Delta R_{xx}$ measured (a) for $H\parallel I$ and (b) for $H\perp I$ and nonreciprocal transverse resistance $\Delta R_{yx}$ (c) for $H\parallel I$ and (d) for $H\perp I$ as a function of electric current ($I$).}
\end{figure}

The $I$ dependence of $\Delta R_{xx}(=\Delta V_{x}/I)$ measured under various magnetic fields applied along the $x$ and $y$ axes is shown in Figs. 2(a) and 2(b). $\Delta R_{xx}$ is only observed for $H||y$ [Fig. 2(b)], and hardly 
observed for $H||-x$ [Fig. 2(a)]. For $H||y$, the magnitude of $\Delta R_{xx}$ is negligibly small without $H$ because of the preserved time-reversal symmetry, and then 
increases with $H$ owing to the enhanced driving force on the vortices and/or the increased vortex density. At higher magnetic fields, superconductivity begins to diminish, and the nonreciprocal response becomes difficult to detect. Note that, in the low-$I$ region, a sign change is observed in the nonreciprocal signal at 1.2 T and 3 T. Similar sign changes have been frequently observed in the nonreciprocal magnetoresistance due to vortex ratchet motion \cite{ideue2020giant,qiao2023electric,du2023superconducting,wu2022nonreciprocal,masuko2022nonreciprocal,zhang2020nonreciprocal,yasuda2019nonreciprocal} and also in former studies on vortex ratchet effects in nanopatterned superconductors with asymmetric pinning potentials \cite{villegas2003superconducting,de2006controlled}. At relatively high magnetic fields, the strong vortex-vortex interaction due to the high vortex density may cause the sign reversal \cite{de2006controlled,lu2007reversible}.

The nonzero $\Delta V_{x}$ signal under $H||y$ indicates that spatial inversion symmetry is broken along the $z$ direction. Lustikova {\it et al.} reported the nonreciprocal magnetoresistance in superconductor/magnet bilayer films originating from the asymmetric surface barriers due to different magnetic environment at interfaces \cite{lustikova2018vortex}. In our case, crystal structure of FeSe is centrosymmetric \cite{hsu2008superconductivity} and thus the symmetry breaking is ascribed to the FeSe/LaAlO$_3$ and FeSe/Si interfaces. The different surface barriers at the FeSe/LaAlO$_3$ and FeSe/Si interfaces break the inversion symmetry along the $z$ axis, satisfying the symmetry requirement for nonreciprocal vortex transport. Note that the breaking of time-reversal symmetry is also required, as the nonreciprocal signal was negligibly small under zero magnetic field~\cite{hashimotofesesuppl}.

We also measured the nonreciprocal transport response along the transverse (Hall) direction. Figures 2(c) and 2(d) show the $I$ dependence of $\Delta R_{yx}$ under $H||-x$ and $H||y$. Nonreciprocal signals are observed for $H||-x$ but negligible
for $H||y$. This is in stark contrast to the case of nonreciprocal magnetoresistance, where the signals appear for $H||y$ but not for $H||-x$, as shown in Figs. 2(a) and 2(b). The overall $I$ dependence of $\Delta R_{yx}$ for $H||-x$ is similar to that of $\Delta R_{xx}$ for $H||y$; $\Delta R_{yx}$ shows a broad dip at a middle $I$ value. The similar $I$ dependences of $\Delta R_{xx}$ and $\Delta R_{yx}$ suggest a vortex origin of the nonreciprocal transverse signal. In terms of the symmetry condition, nonreciprocal transverse signals are expected when $H||I$ for polar systems. Hence, our observation is consistent with the prediction of the symmetry argument. However, we stress that $H$-induced superconducting vortices cannot move when $H||I$ because of the absence of driving forces. In contrast to electronic nonreciprocal transport, where electron flow is always driven by the application of $I$, transport of $H$-induced vortex strongly depends on the mutual directions of $I$ and $H$. Within the framework of the conventional theory of vortex transport, the emergence of the nonreciprocal transverse signal is nontrivial.

\begin{figure}
\includegraphics[width=85mm]{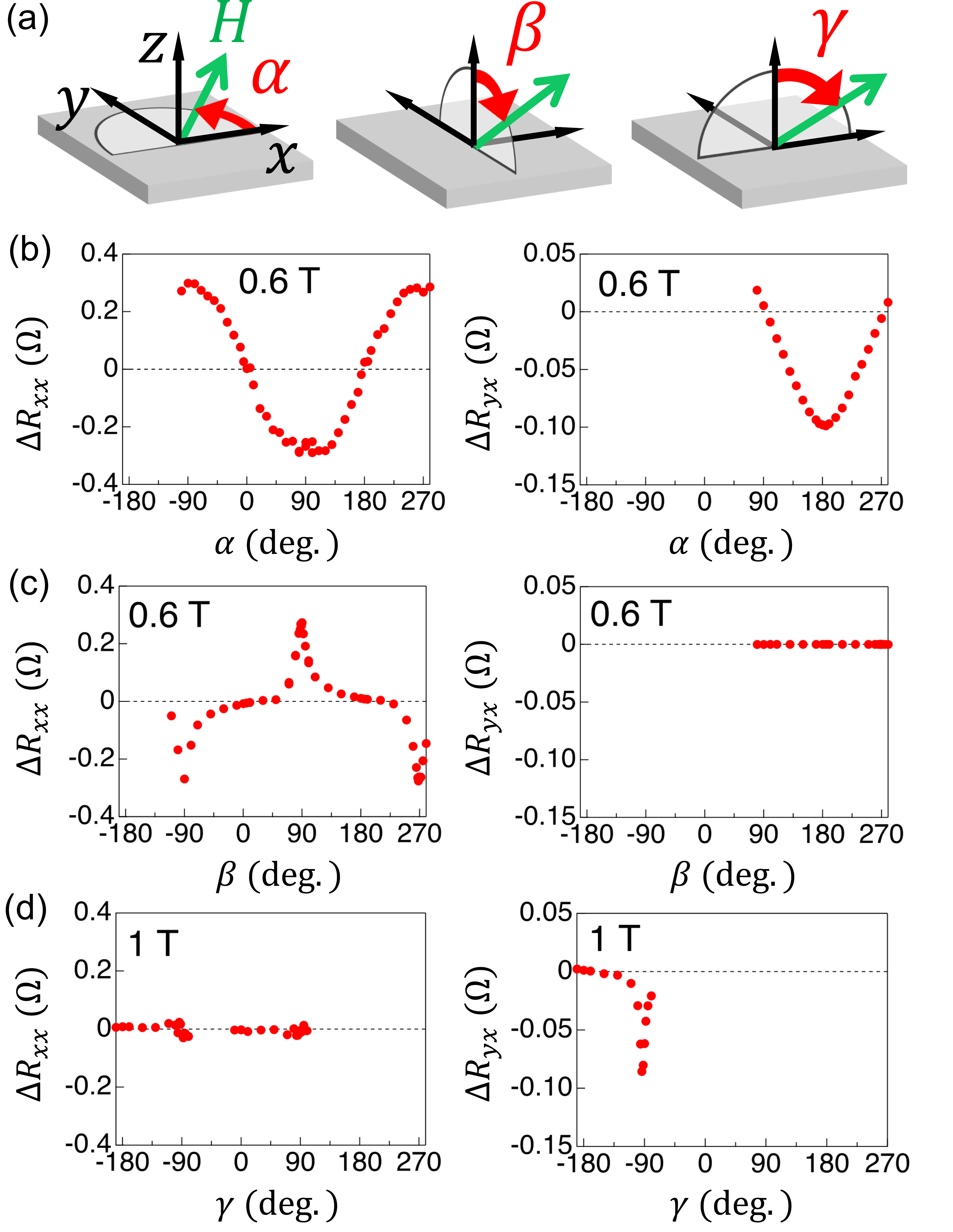}
\caption{\label{fig3} Definitions of magnetic field angles $\alpha$, $\beta$ and $\gamma$. Green arrows represent the direction of the magnetic field ($H$).
(b-d) Magnetic-field angle dependence of $\Delta R_{xx}$ and $\Delta R_{yx}$ in (b) $\alpha$ scan (rotation within the $xy$ plane), (c) $\beta$ scan (rotation within the $yz$ plane), and (d) $\gamma$ scan (rotation within the $xz$ plane). The amplitude of $I$ is 8 mA, and the $H$ strength is 0.6 T for $\alpha$ and $\beta$ scans and 1 T for $\gamma$ scan.}
\end{figure}

In Fig. 3, we measured $\Delta R_{xx}$ and $\Delta R_{yx}$ as a function of magnetic-field angles in three different scan geometries at 2.1 K: $\alpha$-scan in Fig. 3(a), $\beta$-scan in Fig. 3(b), and $\gamma$-scan in Fig. 3(c). The definition of $\alpha$, $\beta$, and $\gamma$ is shown in the upper panels of Fig. 3. 
The results suggest that both $\Delta R_{xx}$ and $\Delta R_{yx}$ satisfy the symmetry condition. In the $\alpha$-scan in Fig. 3(a), $\Delta R_{xx}$ reaches its maximum magnitude at $H||y$ while $\Delta R_{yx}$ is the largest in $H||x$, both of which show an almost sinusoidal angular dependence. Consistently, $\Delta R_{xx}$ and $\Delta R_{yx}$ are barely
observed in the entire $\gamma$ [Fig. 3(c)] or $\beta$ values [Fig. 3(b)], respecetively. As for $\Delta R_{xx}$, similar sinusoidal in-plane angular dependence has been observed in previous reports on nonreciprocal magnetoresistance \cite{masuko2022nonreciprocal,du2023superconducting}. When the magnetic field is rotated from being parallel to the $z$ axis toward the $y$ axis, $\Delta R_{xx}$ remains small up to $\beta \sim 50^{\circ}$ but rapidly increases when the $H$ direction approaches the $y$ direction, exhibiting a peak at $\beta=-90^{\circ},\ 90^{\circ}$, and $270^{\circ}$ in Fig. 3(b). This nonmonotonic angular dependence is likely related to anisotropic pinning effects owing to the layered structure of FeSe \cite{long2008model}. As shown in Fig. 3(c), the angular dependence of $\Delta R_{yx}$ in the $\gamma$-scan is similar to that observed in the $\beta$-scan of $\Delta R_{xx}$. 

The results shown in Fig.~3 suggest that the emergence of $\Delta R_{xx}$ and $\Delta R_{yx}$ is not explained by the vortex Nernst effect \cite{hashimotofesesuppl,pourret2011strong,yang2017bcs,terashima2025}. The absence of $\Delta R_{xx}$ and $\Delta R_{yx}$ in $H||z$ indicates that the vortex Nernst voltage due to the inplane temperature gradient is negligibly small. In the $\alpha$-scan in Fig.~3(a), peak values of $\Delta R_{xx}$ is larger than those of $\Delta R_{yx}$ by a factor of three, which denies that the nonreciprocal signals solely originate from the vortex Nernst effect caused by the out-of-plane temperature gradient: if so, the ratio of them should be equal to that of the sample size along $x$ ($\sim 1.7$ mm) and $y$ ($\sim 1.2$ mm).

\begin{figure}
\includegraphics[width=85mm]{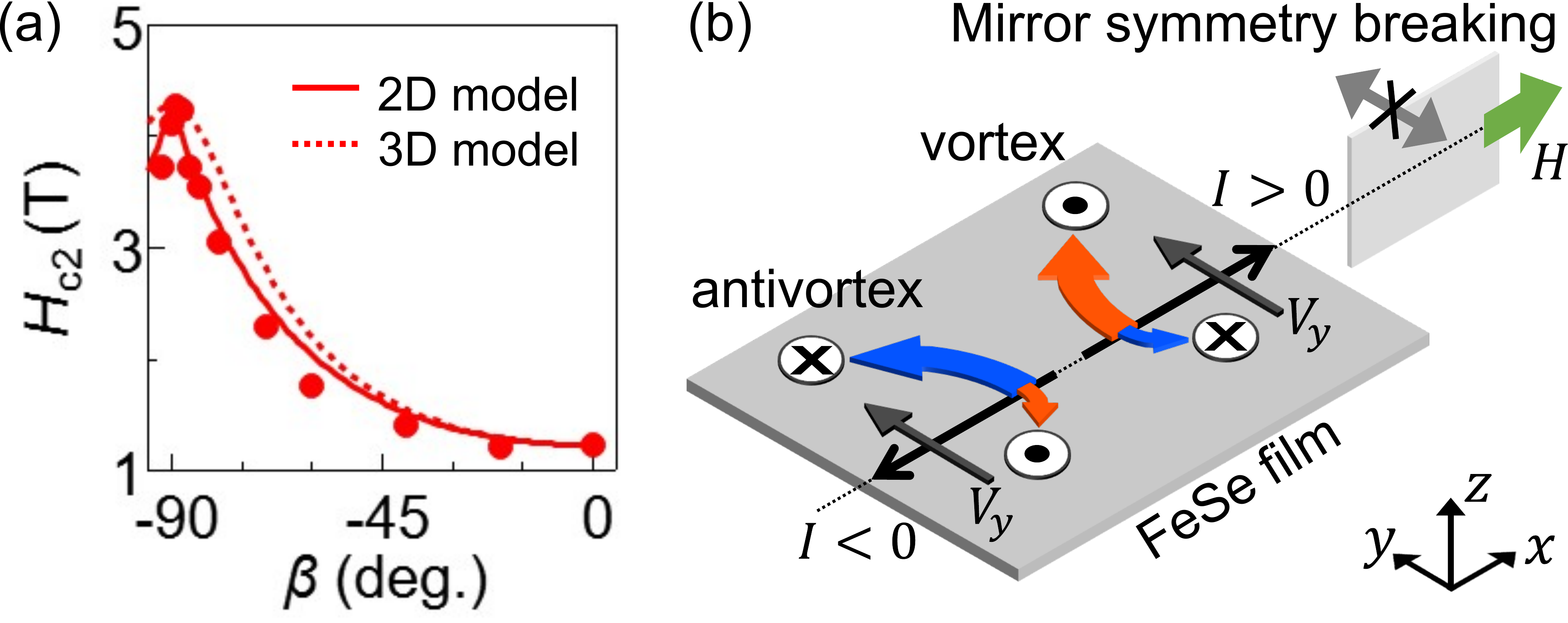}
\caption{\label{fig4} (a) (b) $\beta$ dependence of the upper critical field $H_{\rm{c2}}$ for several $R_{xx}/R_N$ values. Solid and dotted curves stand for the fits to 2D and 3D models, respectively. See SM about the model curves \cite{hashimotofesesuppl}. (b) Schematic illustration of asymmetric transport of thermally excited (anti)vortices when $H||I$. Orange (blue) arrows represent the velocity of vortex (antivortex). }
\end{figure}

Since $H$-induced vortex motion is not expected when $H||I$, the origin of the nonreciprocal transverse response is not straightforward.
A clue to approach the mechanism is a 2D character in superconductivity in FeSe films \cite{farrar2020suppression,zhao2020vortex,schneider2014excess}. We confirmed the 2D nature of our FeSe films by measuring the upper critical field $H_{\rm{c2}}$ at different $\beta$ angles at 2.1 K. $H_{\rm{c2}}$ is evaluated from the $H$ dependence of $R_{xx}$ plotted in Fig.~S1(a) in Supplemental Material (SM) \cite{hashimotofesesuppl}, which is defined as the position at $R_{xx}=0.5R_N$ \cite{farrar2020suppression}. Here $R_N$ is the normal state resistance at 2.1 K and estimated to be $34$ ${\rm \Omega}$. The $\beta$ dependence of $H_{\rm{c2}}$ is plotted in Fig. 4(a). We found that the $\beta$ dependence of $H_{\rm{c2}}$ is better fitted with a 2D model than a three-dimensional model, suggesting that the superconductivity in our FeSe films is 2D in nature. Note that this tendency is robust regardless of the value of $R_{xx}/R_N$ to determine $H_{\rm{c2}}$: see Fig.~S1(b) in SM for the analysis of $H_{\rm{c2}}$ under different $R_{xx}/R_N$ values \cite{hashimotofesesuppl}.

Based on the 2D nature of our FeSe thin film, a plausible origin of the nonreciprocal transverse response is the ratchet motion of (anti)vortices excited without $H$ [Fig. 4(b)]. In 2D superconductors, vortex and anti-vortex pairs are excited thermally or by electric current, in addition to the $H$-induced vortices. The magnetic flux of these (anti)vortices points in the direction perpendicular to the 2D plane, and these (anti)vortices contribute to the nonreciprocal magnetoresistance, as demonstrated theoretically \cite{hoshino2018nonreciprocal} and experimentally in several superconductors \cite{wakatsuki2017nonreciprocal,yasuda2019nonreciprocal,itahashi2020nonreciprocal,masuko2022nonreciprocal,itahashi2022giant,hayashi2024two}. 
Since the direction of the magnetic flux ($||z$) is perpendicular to the $I$ direction ($||x$), the motion of (anti)vortices is allowed along the $y$ axis. 
If the vortex Hall effect is incorporated in the vortex dynamics, it becomes asymmetric for superconductors lacking mirror symmetry with respect to the $xz$ plane \cite{itahashi2022giant,hashimotofesesuppl}.
The asymmetric (anti)vortex transport along the $x$ axis gives rise to nonreciprocal transverse voltage ($||y$), as shown in Fig. 4(b). It is notable that a large vortex Hall angle of $\sim 0.5$ was indeed reported for FeSe single crystals \cite{ogawa2023microwave,okada2021electronic}. 
Although our FeSe film exhibits inversion symmetry breaking only along $z$, the application of $H$ along the $x$ axis further breaks the mirror symmetry with respect to the $xz$ plane, and thus nonreciprocal transverse signal is allowed in terms of symmetry.
We also investigated temperature dependence of the nonreciprocal signals and we find that it is consistent with the scenario of thermally-excited (anti)vortices (see SM about the detail \cite{hashimotofesesuppl}).

We also perform a model calculation for a 2D superconductor under the application of a magnetic field by incorporating the Rashba spin-orbit coupling in the Ginzburg-Landau free energy \cite{hoshino2018nonreciprocal}. As detailed in SM \cite{hashimotofesesuppl}, the vortex velocities calculated by the derivative of the free energy with respect to wave vectors include nonlinear force terms. The structure of the force-velocity relation is similar to that reported for a ratchet potential model in the noncentrosymmetric trigonal superconductor PbTaSe$_2$ \cite{itahashi2022giant}. The nonlinear force terms cause asymmetric transport of (anti)vortices along the $x$ direction in the presence of the vortex Hall effect, leading to the nonreciprocal transverse voltage that is proportional to the Hall angle
under $H||x$, as illustrated in Fig. 4(b). 
This calculation explicitly describes the nonreciprocal transverse signal owing to the mirror symmetry breaking by the magnetic field.
We note that our model calculation also derives the nonreciprocal magnetoresistance in $H||y$ under the same mechanism. In addition, the nonreciprocal magnetoresistance is predicted to be three times larger than the nonreciprocal transverse signal when the width and length of the system are nearly equal, which is
satisfied in this study. Using the peak values of $\Delta R_{xx}$ and $\Delta R_{yx}$ shown in Fig.~2(b) and (c), the ratios of $\Delta R_{xx}$ to $\Delta R_{yx}$ can be estimated as $2 \sim 4$, implying that the asymmetric transport of (anti)vortices mainly contributes to the nonreciprocal signals. We remark that the nonreciprocal motion of $H$-induced vortices can give rise to $\Delta R_{xx}$, but not $\Delta R_{yx}$, which might deviate the value of $\Delta R_{xx}/\Delta R_{yx}$ from the theoretical expectation. See SM about the detail of this scenario \cite{hashimotofesesuppl}.

{\it Conclusion-} 
We successfully observed nonreciprocal charge transport due to vortex ratchet motion in FeSe films not only along the longitudinal direction but also the transverse direction. The nonreciprocal transverse response under $I||H$ is allowed in terms of symmetry, but the microscopic mechanism is nontrivial since vortex motion is not expected in this configuration. The vortex-antivortex pairs thermally excited by the 2D character of FeSe may contribute to the nonreciprocal transverse signal. 
Since this mechanism is not unique to FeSe, we anticipate that similar phenomena will be observed in other 2D superconductors.
The present results highlight rich nonlinear transport phenomena of superconducting vortices.

\begin{acknowledgments}
 The authors are grateful to Dr. T. Kawamura, Dr. S. Sumita, and Prof. Y. Kato for fruitful discussions. This work was supported by JST FOREST Program, Grant Number JPMJFR203H, and by JSPS KAKENHI, Grant Numbers JP23K2244, JP23K26525, JP24H01177, JP24K21726, and JP24K00566.
\end{acknowledgments}

\bibliography{ref_arXiv}

\end{document}